\documentclass[superscriptaddress,
twocolumn,
aps,
prl,
amsmath,amssymb,
reprint,%
]{revtex4-1}

\usepackage{graphicx}
\usepackage{dcolumn}
\usepackage{bm}
\usepackage[utf8]{inputenc}
\usepackage[T1]{fontenc}
\usepackage{mathptmx}
\usepackage{etoolbox}
\makeatother
\begin{document}


\title{The loop-zag resonator: A loop-gap resonator design for improved sensitivity in electron-spin resonance experiments}

\author{Brendan C.~Sheehan}
\affiliation{Department of Physics and Astronomy, Amherst College, Amherst, MA 01002, USA}
\affiliation{Department of Physics, University of Massachusetts Amherst, Amherst, MA 01003, USA}%
\author{Guanchu Chen}
\affiliation{Department of Physics and Astronomy, Amherst College, Amherst, MA 01002, USA}
\affiliation{Department of Physics, University of Massachusetts Amherst, Amherst, MA 01003, USA}%
\author{Sai Chauhan}
\affiliation{Department of Physics and Astronomy, Amherst College, Amherst, MA 01002, USA}
\author{Rilla McKeegan}
\affiliation{Department of Physics and Astronomy, Amherst College, Amherst, MA 01002, USA}
\author{Francisca Abdo Arias}
\affiliation{Department of Physics and Astronomy, Amherst College, Amherst, MA 01002, USA}
\author{William Henshon}
\affiliation{Department of Physics and Astronomy, Amherst College, Amherst, MA 01002, USA}
\author{Charles A.~Collett}
\altaffiliation[Current address: ]{Department of Physics, Hamilton College, Clinton, NY 13323, USA}
\affiliation{Department of Physics and Astronomy, Amherst College, Amherst, MA 01002, USA}
\author{Jonathan R.~Friedman}
\affiliation{Department of Physics and Astronomy, Amherst College, Amherst, MA 01002, USA}
\affiliation{Department of Physics, University of Massachusetts Amherst, Amherst, MA 01003, USA}

\date{\today}

\begin{abstract}
We present a novel design of loop-gap resonator, the loop-zag resonator, for sub-X-band electron-spin resonance spectroscopy. The loop-zag design can achieve improved coupling to small-sample spin systems through the improvement of sample filling factor and RF $B_1$ field. By introducing ``zags'' to the resonator's gap path, the capacitance is increased, accommodating a smaller loop size and thereby a larger filling factor to maintain the requisite resonant frequency. We present experimental spectra on five different resonators, each with approximately the same resonant frequency of $\sim2.9$~GHz, showing that an increase in the number of zags and reduction in loop size gives rise to higher sensitivity. Finite-element simulations of these resonators provide estimates of the improved filling factors obtained through the addition of zags. The frequency range over which this loop-zag design is practical enables a breadth of future applications in microwave engineering, including ESR and ESR-like quantum information microwave techniques.
\end{abstract}

\maketitle

\section{Introduction}
Electron spin resonance (ESR) spectroscopy has long served as a valuable tool across many fields of science and industry; today, several commercial ESR spectrometer models are available on the market. Additionally, many research groups across the world take advantage of nontraditional or home-built ESR hardware. In every ESR spectrometer a method of coupling microwave radiation to the sample is necessary; many spectrometers, including commercial models, use resonant microwave cavities ~\cite{mettAxiallyUniformResonant2001,andersonCavitiesAxiallyUniform2002}. Other spectrometers take advantage of coplanar waveguide (CPW) resonators~\cite{malissaSuperconductingCoplanarWaveguide2013}, microstrip resonators~\cite{johanssonStriplineResonatorESR1974,cebulkaSubKelvin100MK2019}, or loop-gap resonators (LGRs)~\cite{hardySplitRingResonator1981,fronciszLoopgapResonatorNew1982,woodLoopgapResonatorII1984}.

Commercial spectrometers are typically built to operate in a narrow frequency range, often in X-band. At frequencies below X-band, however, cavity resonators often become impractical because increasing radiation wavelength requires cavity sizes that are impractically large. CPW and LGR designs offer the possibility of usage at bands below X-band. CPW designs allow frequency tunability and the use of higher harmonics to work with a wide frequency range, but they often demonstrate poor RF field homogeneity across the sample space~\cite{joshiSeparatingHyperfineSpinorbit2016}. Specialized designs can be used to concentrate the rf magnetic field and enable high filling factor, but do not provide a particularly homogeneous field over the volume of the sample space~\cite{abhyankarScalableMicroresonatorsRoomtemperature2020a,choiUltrastrongMagneticLightmatter2023,ranjanElectronSpinResonance2020}. LGRs, on the other hand, provide a high quality factor ($Q\sim2000$) and good RF field homogeneity, high filling factors, frequency tunability via dielectric insertion into the gap of the resonator, and flexibility of geometry~\cite{twigSensitiveSurfaceLoopgap2010,blankRecentTrendsHigh2017,rinardLoopGapResonators2005}.

The geometry of the LGR is described by the name: each LGR consists of a loop and a gap. LGRs can be described by an equivalent RLC circuit; the loop functions as the inductor and the gap functions as the capacitor. A sample placed within the loop experiences the resonator's RF magnetic field $B_1$, allowing the sample's spectral properties to be probed. With careful adjustment of the position of an antenna, critical coupling with a $Q\sim2000$ is typically achieved. Because the inductance $L$ and capacitance $C$ depend on the geometry of the loop and gap, respectively, the resonant frequency $\omega=(LC)^{-1/2}$ can be tuned by adjusting the thickness of the resonator, the width and length of the gap, and/or the radius of the loop. Furthermore, insertion of a dielectric material into the gap provides an additional method of tuning the capacitance and therefore the resonant frequency.

The filling factor $\eta$ can be written as
\begin{equation}
\eta = \frac{\int_\text{sample}|B_1|^2dV}{\int_\text{mode}|B_1|^2dV};
\label{filleq}
\end{equation}
this describes the ratio of energy felt by the sample to the total energy in the resonant mode, typically the volume of a shielded space containing the resonator. The continuous wave ESR signal size $\nu$ is related to the filling factor by $\nu = \chi Q_L\eta P_0^{1/2}$, where $\chi$ is the RF susceptibility of the sample, $Q_L$ is the loaded $Q$, and $P_0$ is the incident power~\cite{feherSensitivityConsiderationsMicrowave1957,hydeMultipurposeLoopgapResonator1989}. The sample response in the nonsaturating regime therefore increases linearly with $\eta$.

One way to increase $\eta$ is by maximizing the amount of sample in the loop, but there are many cases in which obtaining larger samples is impractical or impossible. In such cases, an improved resonator design that increases filling factor by more effectively focusing the $B_1$ field onto the sample can be used. One method of achieving such an effect is to reduce the loop radius to improve the filling factor; this results in a smaller resonator inductance that may shift the resonant frequency to an undesirably high range. Thus, there are contexts in which a traditional LGR cannot properly provide the desired resonant frequency while furnishing a high filling factor and a strong RF magnetic field. In this paper we introduce an LGR design that addresses this conundrum: the loop-zag resonator (LZR)~\cite{friedmanLoopGapResonators2021}. The LZR design takes advantage of similar physics as the LGR; however, the gap can be continuously elongated through the use of ``zags'' in the gap path: the gap can wind back and forth across the plane of the resonator, producing in an interdigitated capacitor. The resulting increased capacitance allows one to maintain the desired resonant frequency while working with a small loop radius. The smaller loop yields both a stronger RF magnetic field for a given input RF power, and an increased filling factor.

\begin{figure}
    \centering
    \includegraphics[width = 0.48\textwidth]{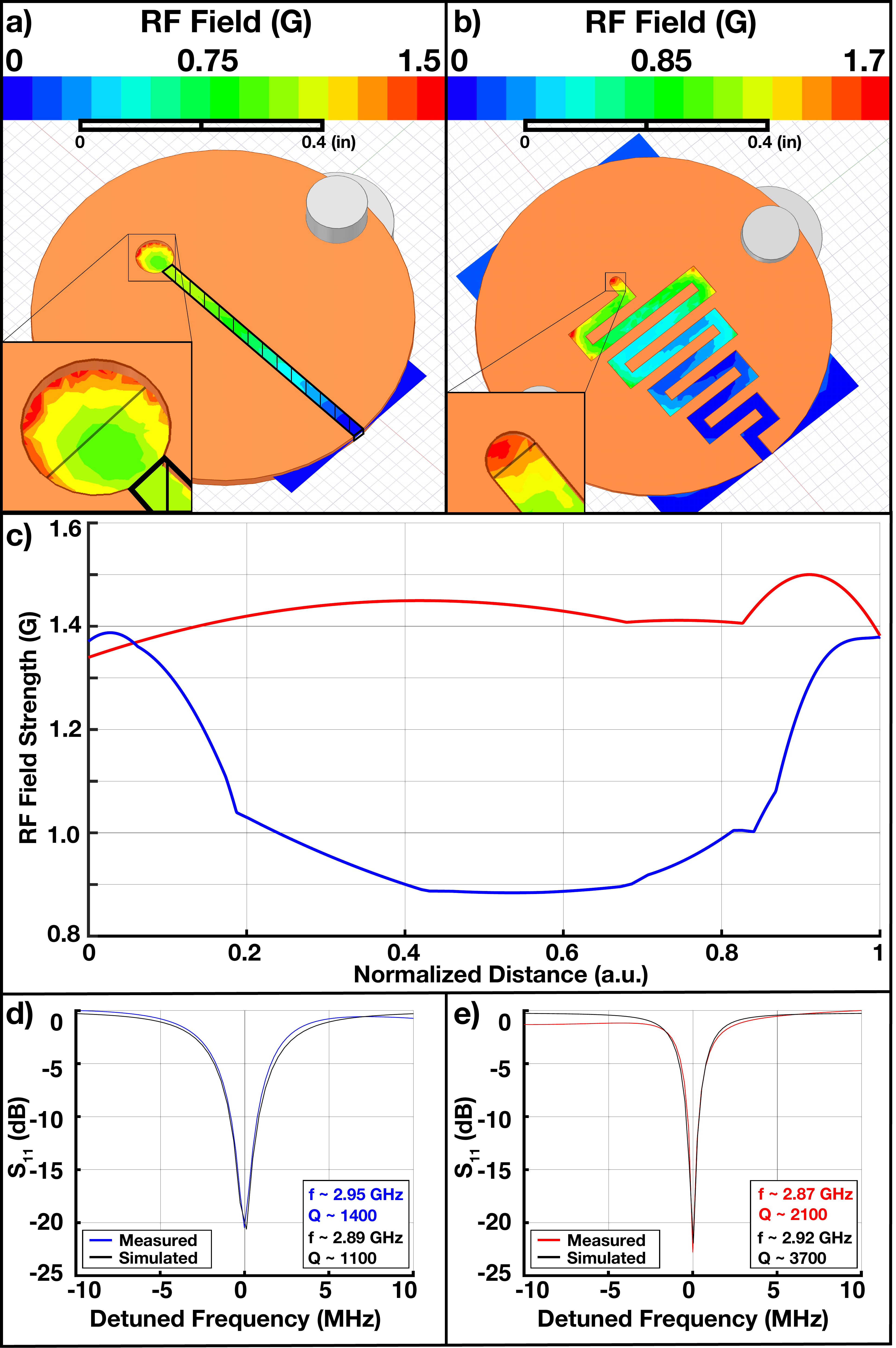}
    \caption{Loop-gap resonator (LGR) and four-zag loop-zag resonator (LZR) designs are shown in panels (a) and (b), respectively. The colormap in each figure shows the strength of the RF ($B_1$) field in each resonator, for an RF input power of 1 mW (0.0 dBm). Note each colormap has a different scale; the LZR shows a higher maximum field in the loop than the LGR. The black framed outline in the LGR gap in (a) shows placement of dielectric material (rutile). Insets show focus on loops of each resonator; shaded black lines cutting diagonally across each loop in the inset provide references for which RF field strength is shown in (c): RF field as a function of distance from the left edge of the loop. Traces shown in (c) go through the center of the loop. (d) and (e) show comparisons of simulated and measured response of the LGR and four-zag LZR as a function of detuning, respectively. Measured $Q$ and resonant frequencies are given in blue (d) and red (e), while simulated $Q$ and frequencies are given in black for both panels. All simulations were produced using Ansys Electronics Desktop Suite, release 2022 R1.}
    \label{resonators}
\end{figure}

The LZR design could permit strong coupling between the a small number of spins and the resonator, potentially opening up a number of novel applications in quantum information, such as using single spins as qubits or using spin systems as a form of quantum memory~\cite{grezesStorageRetrievalMicrowave2015,wuStorageMultipleCoherent2010}. ``Strong coupling'' of the radiation to a collection of $N$ spins requires the vacuum Rabi frequency $\Omega^N_{VR}=\sqrt{N}g\mu_B B^{(1)}_1/\hbar$ to be greater than the relaxation times of the resonator and the spins~\cite{kuboStrongCouplingSpin2010,schusterHighCooperativityCouplingElectronSpin2010,chiorescuMagneticStrongCoupling2010}. Here $B^{(1)}_1$ is the RF field of a single photon, $g$ is the Land\'e $g$-factor, and $\mu_B$ is the Bohr magneton. In practice, this requirement results in $\Omega^N_{VR}/2\pi > \sim1-10$~MHz when working with solid-state spins. Unlike cavity and CPW resonator designs, the LGR/LZR are lumped-element devices (to a first approximation) and therefore the field distribution is not constrained by the Rayleigh limit, allowing for miniaturization of the loop and concentration of the field within the loop volume. If the RF field does not substantially leak outside of the loop, $B^{(1)}_1$ scales with loop volume $V$ as $V^{-1/2}$ while the number of spins coupled to the loop scales as $N\propto V$, so by scaling the loop down to the nanoscale one can in principle couple to a small number of spins or a single spin while remaining in the strong coupling regime. The LZR design thus enables an important step toward the single-spin regime through minimization of the loop size.

\section{Design}

For this study, we designed six resonators using the Ansys Electronics Desktop Suite and fabricated five of the designs out of oxygen-free high thermal conductivity (OFHC) copper using a CNC milling machine -- see Table~\ref{table}. One resonator, the five-zag LZR, had a gap width/loop radius that was too small to be machined with the available equipment.  Two of our resonators are shown in Fig.~\ref{resonators}. The standard LGR design, shown in Fig.~\ref{resonators}(a), features a long gap with a gap width of 0.02~in, a gap length of 0.42~in, and a loop radius of 0.03~in. The thickness of the resonator is 0.02~in. Furthermore, the use of a dielectric brings the simulated resonant frequency in Fig.~\ref{resonators}(a) from 4.38~GHz to 2.89~GHz. The dielectric is represented by a black-edged block with diagonal hashes filling the length of the gap. The four-zag LZR (Fig.~\ref{resonators}(b)) has similar features; the gap width is 0.021~in. and resonator thickness remains 0.02~in. The loop radius has been reduced to 0.0105~in, making the loop a semicircle that terminates the gap. The increased gap length of the LZR allows for a smaller loop while yielding a resonant frequency of 2.92 GHz, similar to the LGR design with dielectric.

The insets in Figs.~\ref{resonators}(a) and \ref{resonators}(b) zoom in on each resonator's loop to show the simulated $B_1$ field strength distribution in the loop. To perform field simulations, input power was set to 0.0~dBm. While quality factor was not fixed, it was relatively consistent ($\sim2000$) among the five designs (Table~\ref{table}).

Each inset shows a black line across the diameter of the loop at the midpoint of the resonator thickness. Fig.~\ref{resonators}(c) shows the $B_1$ field strength calculated along each of these lines.  The LGR's large loop size causes a decrease in $B_1$ field strength at the center of the loop (shown in blue); the sample can therefore either be placed at the center of the loop (to take advantage of the more homogeneous $B_1$ field) or near the edge (to maximize the field strength). The LZR, however, allows the sample to experience a large, relatively homogeneous $B_1$ field throughout the loop (Fig.~\ref{resonators}(c), shown in red). The $B_1$ field strength in the center of the four-zag LZR loop is $\sim65\%$ larger than the field strength in the center of the LGR loop.  Averaging over the loop volume, we find the mean field strength is 26\% larger in the four-zag LZR compared to the LGR. 

\begin{table*}[ht!]
\centering
\begin{tabular}{||c|c|c|c|c|c|c||} 
 \hline
 \textbf{Design} & \textbf{Resonant Frequency} & \textbf{Quality Factor} & \textbf{Loop Radius} & \textbf{Gap Length} & \textbf{Gap Width} & \textbf{Mean Field in Loop} \\ 
 \hline\hline
 \textbf{LGR} & 2888 MHz & 1400 & 0.03 in & 0.42 in & 0.02 in &  1.061 G \\
 \hline
 \textbf{One-zag LZR} & 2887 MHz & 2200 & 0.01 in & 1.16 in & 0.02 in & 1.295 G \\
 \hline
 \textbf{Two-zag LZR} & 2956 MHz & 2300 & 0.0225 in & 1.25 in & 0.021 in & 0.914 G \\
 \hline
 \textbf{Three-zag LZR} & 2976 MHz & 2900 & 0.015 in & 1.49 in & 0.021 in & 1.136 G \\
 \hline
 \textbf{Four-zag LZR} & 2921 MHz & 3700 & 0.0105 in & 1.68 in & 0.021 in & 1.339 G \\
 \hline
 \textbf{Five-zag LZR} & 2959 MHz & 3100 & 0.0025 in & 1.165 in & 0.005 in & 3.972 G \\
 \hline
\end{tabular}
\caption{Parameter values for each resonator in simulations performed for Figs.~\ref{resonators} and~\ref{filling}. The resonant frequency of each resonator could be tuned by changing the loop radius, the gap length, and the gap width. The resonator thickness was 0.02" for all resonators. The input power was set identically in simulation for each, to 0.0~dBm. The antenna position was adjusted to achieve a coupling of -20~dB, undercoupled, for each design; the quality factor was determined when this condition was met. The LGR's quality factor is diminished due to the inclusion of the rutile dielectric with loss tangent $\delta=0.001$.}
\label{table}
\end{table*}

To calculate both the $B_1$ field strength and the filling factor, several parameters need to be held constant for the sake of comparison between resonator designs. The filling factor is referenced to the total magnetic field energy in the system $E_\text{B} = \frac{1}{2\mu_0}\int_\text{mode} B_1^2dV = \frac{P_0 Q}{2\omega}$, where $P_0$ is the incident power and $\omega$ is the resonant frequency. Parameters for each resonator simulation are shown in Table~\ref{table}.

Figures~\ref{resonators}(d) and (e) show a comparison of the simulated and measured resonances. Coupling was set to $-20$~dB in our experiments since working with a slightly undercoupled resonance appears to improve sensitivity. The same undercoupling condition was implemented in the simulations. Each panel notes the $Q$ value measured for each resonator. We measured the LGR $Q$ value to be approximately half that of the LZR $Q$ due to the use of the dielectric material (rutile); in simulation we determined the dielectric loss tangent for the rutile to be $\delta=.001$ via comparison with the measured resonance.

The introduction of an interdigitated capacitor via the addition of zags increases the capacitance by increasing the length of the gap. In addition to a strong RF field in the loop, Fig.~\ref{resonators}(b) shows some RF field extending into parts of the gap, indicating some parasitic inductance of the capacitor.  To quantify the enhanced sensitivity of the LZR while accounting for the extraneous RF field outside the loop, we calculate the filling factor $\eta$ as per Eq.~\ref{filleq},
where the volume of the mode is considered to be the volume of the entire shielded space in which the resonator sits, and the volume of the sample is taken to be a small spherical volume of radius 0.0025" near the center of the loop that approximates the actual volume of the sample used in experiments. The results of these calculations are shown in Fig.~\ref{filling}, overlaid with the results of sample measurement discussed below.

\section{Results}

To compare the sensitivity of each resonator, we measured an ESR standard sample, \textit{2,2-Diphenyl-1-picrylhydrazyl} (DPPH). DPPH (C$_{18}$H$_{12}$N$_5$O$_6$) is an organic radical with a $g=2.0036$ ESR signal coming from an unpaired electron in each molecule in the ensemble. It is commercially available as a powder sample and has been used as an ESR standard for decades~\cite{slifkinUseDPPHStandard1975,yordanovDPPHPrimaryStandard1994}. Inhomogeneous broadening results in  a full width at half maximum (FWHM) typically measured between 1 and 5~Oe at ESR frequencies below Q-band. These characteristics make it an good continuous wave (cw) ESR standard, but its short relaxation times are such that pulsed ESR techniques cannot be used to extract a Hahn echo response. 

We employed a heterodyne cw technique with a homebuilt ESR spectrometer built from commercially available microwave components and a Tabor SE5082 programmable arbitrary waveform generator (see Appendix for details). We experimentally examined four LZR designs, each with a measured resonant frequency between 2.89~GHz and 3.05~GHz. A typical measured quality factor for all four LZR designs was $Q\sim2500$ at 10~K. The LGR has a resonant frequency of 4.38~GHz and $Q\sim2500$ at 10~K. Similar to the simulations performed for Fig.~\ref{resonators}, we inserted a piece of rutile, with dielectric constant of 86, into the gap of the LGR to bring its resonant frequency down to 2.95~GHz (very close to the simulated resonant frequency of 2.89~GHz). The large dielectric loss tangent of rutile pushed the LGR $Q$ to $Q\sim1100$ at 10~K. The sample and resonator were placed in an OFHC copper shield at the bottom of a custom-designed ESR probe~\cite{joshiAdjustableCouplingSitu2020}.  To avoid power broadening and saturation, the ESR cw power was reduced until the FWHM of the measured signal peak no longer varied. 

We used the same sample of DPPH to characterize each resonator. By encasing the DPPH powder in a thin layer of vacuum grease, the sample could be transferred between resonators with no loss of powder. We estimated the sample to contain $\sim7\times10^{15}$ spins within the resonator's loop~\cite{williamsCrystalStructure2diphenyl1picrylhydrazyl1967}.

\begin{figure}[ht!]
    \centering
    \includegraphics[width = 0.48\textwidth]{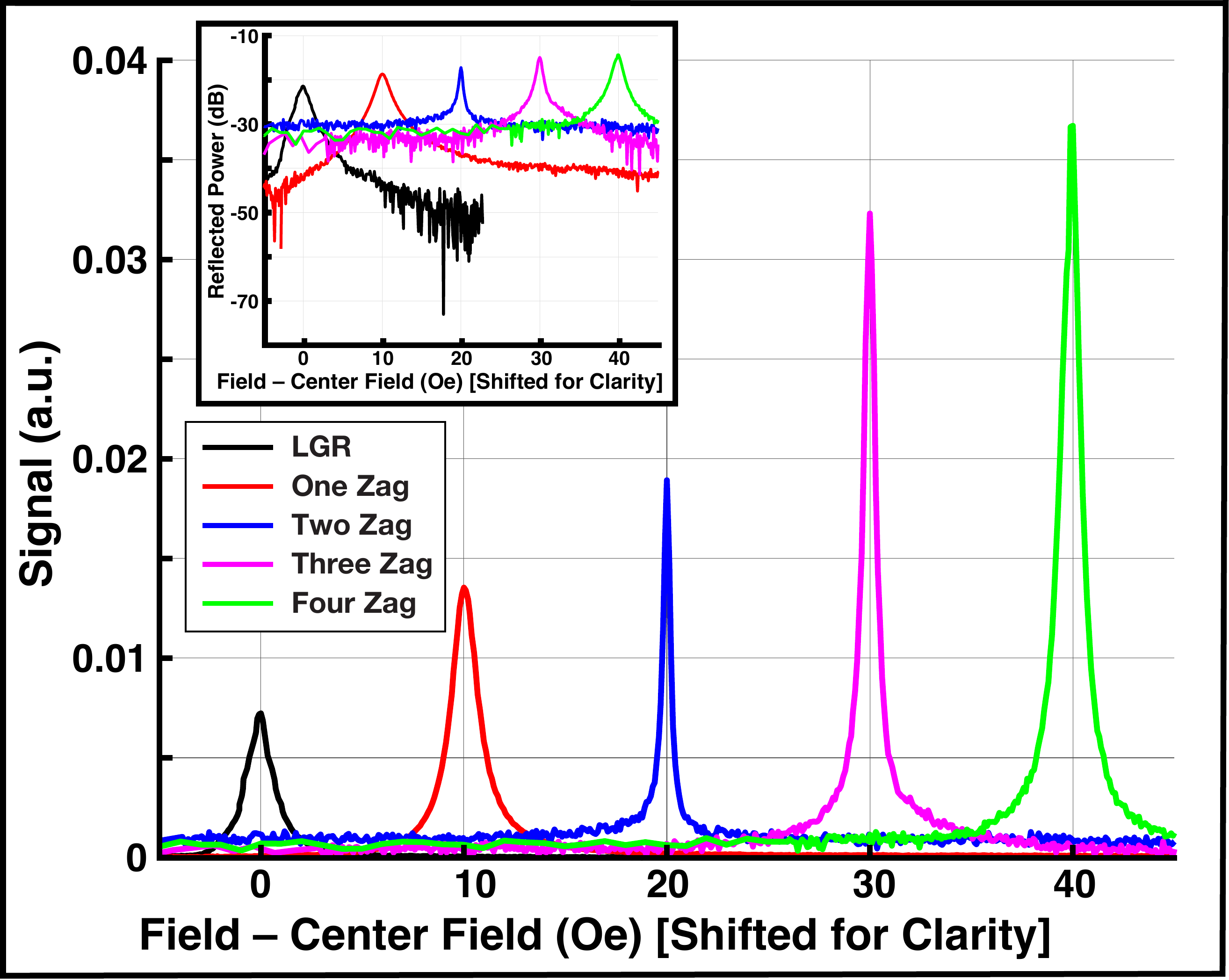}
    \caption{CW ESR measurement of DPPH at the resonant frequency of each resonator at 10~K. For the LGR and the one-zag LZR, the cw power applied to the sample was $-45$~dBm. For the two, three, and four-zag resonators, the cw power applied was $-55$~dBm. Signal size increased as the number of zags increased, with the largest signal from the four-zag resonator. The inset shows a semilog scale of the same data.}
    \label{data}
\end{figure}

Figure \ref{data} shows spectra taken of the DPPH powder sample in each resonator at 10~K. To quantify the signal, we integrate the observed ESR peak with respect to magnetic field.  Integrated signal strength from DPPH in the four-zag LZR (green) was 5.4 times larger than the integrated signal strength from the LGR (blue) at comparable frequencies. The measured FWHM for DPPH from each resonator ranged from was 1.63~Oe (the one-zag resonator, red) to 0.52~Oe (the two-zag resonator, black), in agreement with previously established values~\cite{chaudhuriDetectionLbandElectron2020}. The inset of Fig.~\ref{data} shows a semilogarithmic scale of the same data, making the background noise of the spectra more visible. 

\begin{figure}[ht!]
    \centering
    \includegraphics[width = 0.48\textwidth]{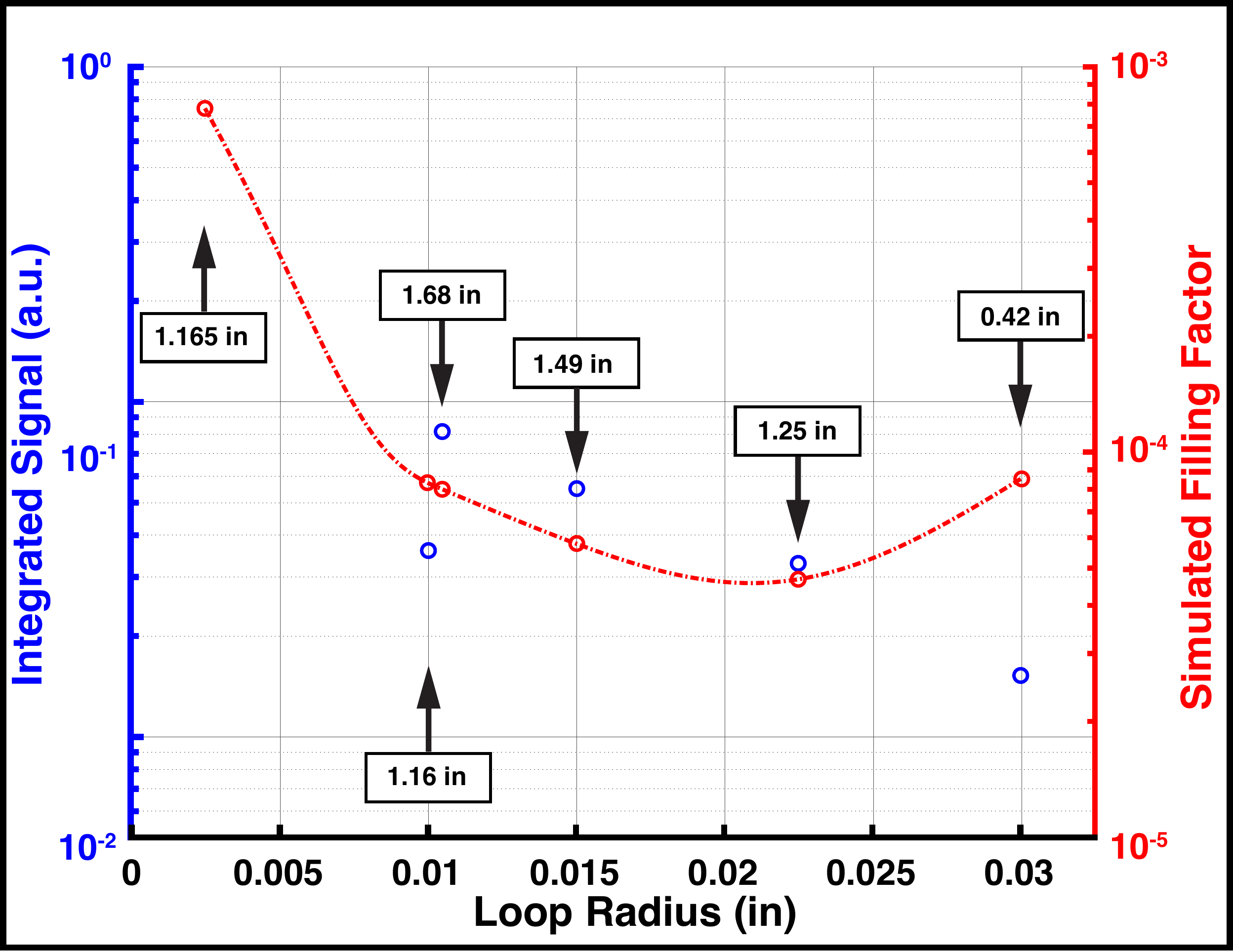}
    \caption{Blue shows the integrated signal strength, integrated with respect to magnetic field from Fig.~\ref{data}, as a function of loop radius; red shows the calculated filling factor. The red dashed curve is a guide to the eye.  Boxes show the length of the gap for each resonator design. The LGR and four-zag LZR (shown in Figs.~\ref{resonators}(a) and~\ref{resonators}(b), respectively) have gap lengths of 0.42~in. and 1.68~in, respectively. For most resonators, the integrated signal size scales with the filling factor, indicating that the minimization of the loop radius allows for a better field concentration on the sample and therefore a stronger sample response. All simulations were produced using Ansys Electronics Desktop Suite, release 2022 R1.}
    \label{filling}
\end{figure}

Figure~\ref{filling} shows the calculated filling factor (red) as a function of loop radius for the six resonator designs.  This shows a clear increase in filling factor as the loop radius is reduced. The largest calculated filling factor was found for the five-zag LZR, which has the smallest loop radius.  While that design was not tested, the signal from the five experimental resonators (blue) from DPPH shows a similar trend to the calculations. Introduction of zags in the resonator gap affords a smaller loop radius and therefore improved signal size.

\section{Discussion}

In quantum information, one generally needs to control and measure the quantum state of individual qubits. Many qubit experiments, such as those using superconducting qubits, are done at microwave frequencies of a few GHz, often employing 1D (stripline, CPW) or 3D (cavity) geometries. Such resonators work well with superconducting qubits, which have large transition dipole moments. Potential nanoscale qubits such as phosphorus~\cite{mortonSolidstateQuantumMemory2008} or bismuth~\cite{mortonStoringQuantumInformation2018} impurities in silicon, or molecular nanomagnets~\cite{shiddiqEnhancingCoherenceMolecular2016}, however, have smaller moments and ESR signals are found in bulk but rarely on the single-spin level. To see single-spin ESR requires specialized techniques like STM~\cite{rugarSingleSpinDetection2004}, ODMR~\cite{grotzSensingExternalSpins2011}, or EDMR~\cite{harneitRoomTemperatureElectrical2007}. The LZR approach potentially allows for miniaturization to the scale of single qubits, both superconducting and even individual spin-based systems, and would afford addressability of single qubits. In the strong coupling regime, photon states could be ``parked'' in the qubits and later retrieved. Many experiments on superconducting qubits are done near zero field as a field would disrupt superconductivity. For such applications, an LZR could be fabricated out of superconducting metals such as niobium or aluminum instead of the Cu used here for ESR. A superconducting LZR would have a significantly higher $Q$ than one made of copper. 

For our five-zag design, we estimate a single-photon magnetic field of $B^{(1)}_1=43$~pT. This is comparable to or larger than the field strengths found in mode-engineered cavity designs~\cite{choiUltrastrongMagneticLightmatter2023}. This demonstrates the primary utility of the LZR design: it affords high RF field strengths, which translates into a large filling factor in ESR experiments, has large $Q$, and, importantly maintains a good field homogeneity over the working volume of the resonator. By increasing the number of zags and incorporating the use of a low-loss dielectric, the resonator capacitance can be enhanced further, potentially allowing the loop size to be reduced substantially. This, in turn, would result in larger RF magnetic fields in the loop, a greater filling factor, and the ability to measure, manipulate and control systems containing small numbers of spins.

In conclusion, we have presented a novel loop-gap resonator design for the purpose of coupling to smaller samples while preserving the desired resonant frequency. This loop-zag resonator design takes advantage of increased capacitance to allow for a smaller loop with a more concentrated $B_1$ field coupling to the sample, resulting in a larger filling factor. Characterization of an ESR standard sample shows increased signal with smaller loop radius and larger number of zags. This LZR design represents an important step toward the ability to couple to and address systems with a limited number of spins.

\begin{acknowledgments}
We thank G.~Joshi and D.~S.~Hall for useful conversations and advice, and J.~Kubasek for assistance with the machining of the resonators. Support for this work was provided by the National Science Foundation under grant nos.~DMR-1708692 and DMR-2207624.
\end{acknowledgments}

\appendix*
\section{Appendix}

\begin{figure}[ht!]
    \centering
    \includegraphics[width = 0.48\textwidth]{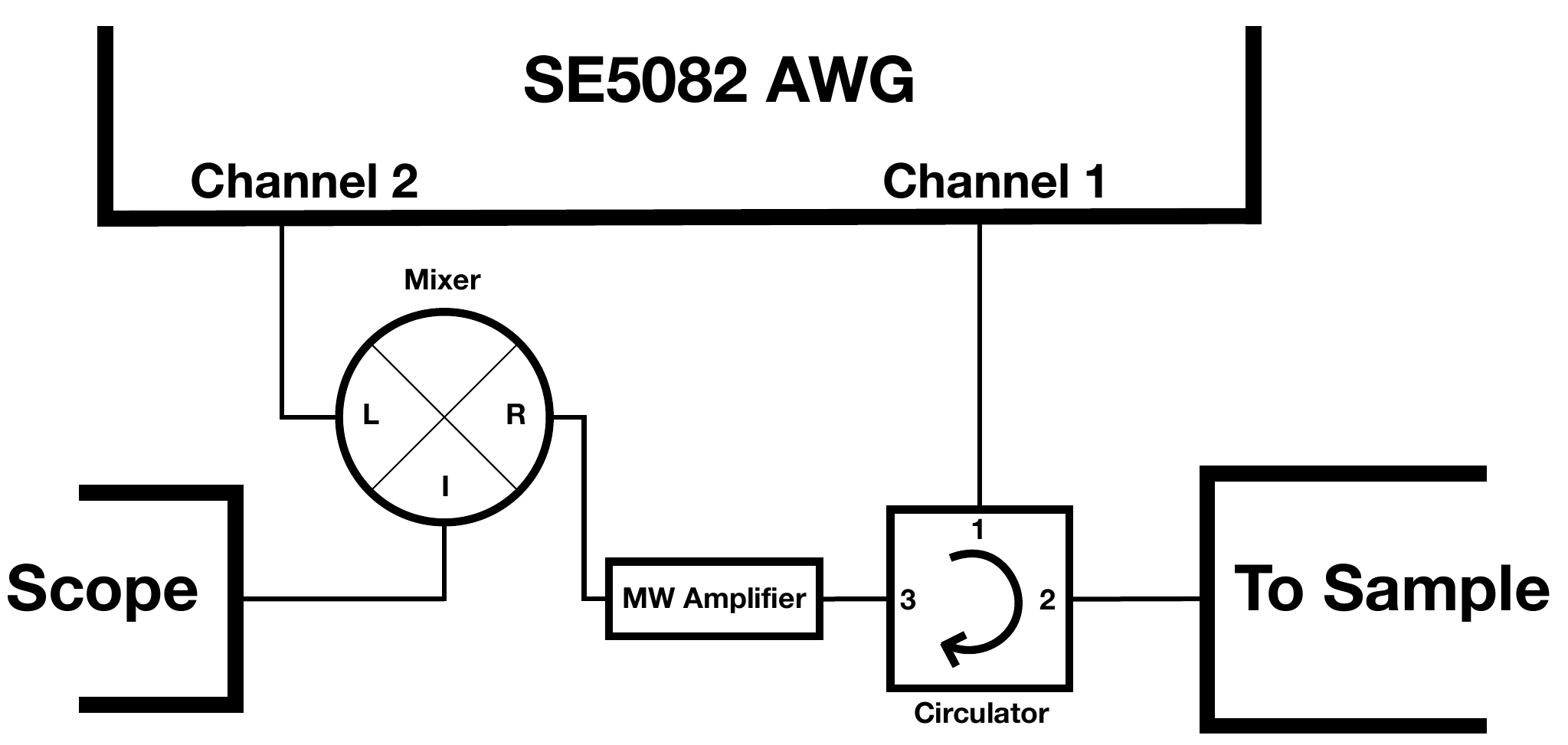}
    \caption{Block diagram showing the spectrometer used for direct-detection ESR. Channel 1 of the AWG produced the ESR working frequency while Channel 2 produced the reference modulated from the working frequency by 1~MHz. All components in the circuit, aside from the resonators, are commercially available.}
    \label{diagram}
\end{figure}

The home-built spectrometer designed to perform the experiments presented in this paper was developed from commercially available microwave components and a Tabor Electronics SE5082 arbitrary waveform generator (AWG). A block diagram depicting the circuit is shown in Fig.~\ref{diagram}. Channel 1 of the AWG produces the cw ESR signal, which is sent to the sample via a coaxial line fed into ESR probe~\cite{joshiAdjustableCouplingSitu2020}. The custom-built probe is built for direct-detection reflection spectroscopy; the resonator coupling can be tuned \textit{in situ}. For all experiments in this paper, the microwave radiation is slightly undercoupled (to -20 dB loss below coaxial background at resonance). The probe is integrated into a Quantum Design Physical Property Measurement System (PPMS) cryostat with a base temperature of 1.8~K. A circulator sends the return signal to a microwave mixer, which mixes the signal with a reference from channel 2 of the AWG. The reference is offset from the ESR frequency by 1~MHz. The 1~MHz down-converted output of the mixer is measured with a Tektronix mixed-domain oscilloscope and read by software for analysis. A virtual lock-in technique is implemented with software, whereby the signal is demodulated, filtered with a low-pass second-order Butterworth filter, and integrated.

\bibliographystyle{apsrev4-1}
\bibliography{LGR_LZR_bibtek}

\end{document}